\documentclass[a4paper,
              ]{jacow}

\makeatletter%
	\ifboolexpr{bool{xetex}}
	 {\renewcommand{\Gin@extensions}{.pdf,%
	                    .png,.jpg,.bmp,.pict,.tif,.psd,.mac,.sga,.tga,.gif,%
	                    .eps,.ps,%
	                    }}{}
\makeatother

 {\usepackage[utf8]{inputenc}}           

\ifboolexpr{bool{xetex} or bool{luatex}} 
 {}                                      

\usepackage[USenglish]{babel}			 
\usepackage[final]{pdfpages}
\usepackage{multirow}
\usepackage{ragged2e}
\usepackage{placeins}
\usepackage{subcaption}
\usepackage{cite}
\ifboolexpr{bool{jacowbiblatex}}%
 {%
  \addbibresource{jacow-test.bib}
  \addbibresource{biblatex-examples.bib}
 }{}
\begin{document}

\title{Proton Beam Defocusing as a Result of Self-Modulation in Plasma}

\author{Marlene Turner,\textsuperscript{1}\thanks{marlene.turner@cern.ch} CERN, Geneva, Switzerland\\
Alexey Petrenko, Edda Gschwendtner, CERN, Geneva, Switzerland\\
Konstantin Lotov,\textsuperscript{2} Alexander Sosedkin,\textsuperscript{2} Novosibirsk State University, 630090, Novosibirsk, Russia\\
\textsuperscript{1}also at Technical University of Graz, Graz, Austria\\
\textsuperscript{2}also at Budker Institute of Nuclear Physics SB RAS, 630090, Novosibirsk, Russia}
	
\maketitle

\begin{abstract}
The AWAKE experiment will use a \SI{400}{GeV/c} proton beam with a longitudinal bunch length of $\sigma_z = 12\,\rm{cm}$ to create and sustain GV/m plasma wakefields over 10 meters \cite{AWAKE}. A \SI{12}{cm} long bunch can only drive strong wakefields in a plasma with $n_{pe} = 7 \times 10^{14}\,\rm{electrons/cm}^3$ after the self-modulation instability (SMI) developed and microbunches formed, spaced at the plasma wavelength. The fields present during SMI focus and defocus the protons in the transverse plane \cite{SMI}. We show that by inserting two imaging screens downstream the plasma, we can measure the maximum defocusing angle of the defocused protons for plasma densities above $n_{pe} = 5 \times 10^{14}\,\rm{electrons/cm}^{-3}$. Measuring maximum defocusing angles around \SI{1}{mrad} indirectly proves that SMI developed successfully and that GV/m plasma wakefields were created \cite{Turner2016}. In this paper we present numerical studies on how and when the wakefields defocus protons in plasma, the expected measurement results of the two screen diagnostics and the physics we can deduce from it.
\end{abstract}

\section{INTRODUCTION}
The Advanced Proton-Driven Plasma Wakefield Acceleration Experiment (AWAKE) is currently under construction at CERN. AWAKE will use a \SI{400}{GeV/c} proton beam from the CERN SPS to excite GV/m plasma wakefields. The goal of AWAKE is to accelerate electrons to GeV energies in \SI{10}{m} of plasma \cite{AWAKE}. 

AWAKE uses a \SI{10}{m}-long rubidium vapour source with a density of $7 \times 10^{14}\,\rm{electrons/cm}^3$ \cite{PLASMA}. A \SI{450}{mJ} laser  (4 TW, \SI{100}{fs}) \cite{LASER} ionizes the outermost electron of the rubidium atom and creates a plasma with a radius of \SI{1}{mm}. The \SI{400}{GeV/c} proton drive bunch has a longitudinal bunch length of $\sigma_z = 12\,\rm{cm}$, a radial bunch size $\sigma_r = 0.2\,\rm{mm}$, $3\times 10^{11}$ protons per bunch and an emittance of \SI{3.6}{mm mrad}. 

Using the linear plasma wakefield theory and the condition of most efficient wakefield excitation: $\sigma_z = \sqrt{2} c/\omega_{pe}$, where $c$ is the speed of light, $\omega_{pe} = \sqrt{4 \pi n_{pe} e^2/m_e}$ the plasma electron frequency, $n_{pe}$ the plasma electron density, $e$ the electron charge and $m_e$ the electron mass, we can estimate that the optimum plasma density for $\sigma_z = 12\,\rm{cm}$ is $n_{pe} = 4 \times 10^{9}\,\rm{electrons/cm}^3$, which corresponds to a maximum accelerating field of $\approx 6\,\rm{MV/m}$.
To create GV/m plasma wakefields AWAKE will use a plasma density of $n_{pe} = 7 \times 10^{14}\,\rm{electrons/cm}^3$ which corresponds to an optimum drive bunch length of $\sigma_z \approx 0.3\,\rm{mm}$.
  
The energy of the proton bunches available at CERN is enough to accelerate electrons up to the TeV range \cite{PATHTO}. To reach GV/m wakefields, the experiment must rely on the self-modulation instability (SMI) to modulate the long proton bunch into micro-bunches spaced at the plasma wavelength $\lambda_{pe}$ ($\lambda_{pe} = 1.2\,\rm{mm}$ for $n_{pe} = 7 \times 10^{14}\,\rm{electrons/cm}^3$). These micro-bunches can then resonantly drive plasma wakefields.

The SMI is seeded by the ionization front created by a short laser pulse overlapping with the proton bunch \cite{PATHTO}. The transverse wakefields periodically focus and defocus protons, depending on their position $\xi$ along the bunch; the proton bunch, while being modulated, contributes to the wakefield growth.

Inserting two imaging screens downstream of the plasma, gives the possibility to detect the protons that are focused and defocused by the SMI.
 
Phase 1 of the AWAKE experiment will start in late 2016. The goal is to measure and understand the development of the SMI. In AWAKE phase 2 (2017-2018) externally injected electrons with the energy $\approx\,$ \SI{15}{MeV} will be accelerated to GeV energies.

In this paper, we study the wakefields that act on the defocused protons and we look at the proton trajectories. We describe the measurable quantities of the two-screen images and show how we plan to determine the saturation point of the SMI from these images.
  
\section{Numerical Simulations}
\subsection{Wakefields acting on the proton drive bunch}

Plasma simulations were performed with the quasi-static 2D3v code LCODE \cite{LCODE1,LCODE2}. All following simulations use the AWAKE baseline parameters described in the Introduction section except that we used a more realistic proton beam emittance of $2.2\,\rm{mm}\,\rm{mrad}$ instead of $3.6\,\rm{mm}\,\rm{mrad}$ \cite{EMITT}.

\begin{figure}[htb!]
		\includegraphics[width = 0.5\textwidth]{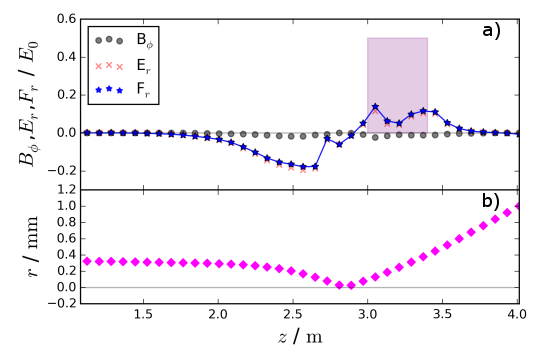}
		\caption{(a) Fields and the transverse force acting on the maximum defocused proton as a function of the longitudinal position $z$ along the plasma. (b) The radial proton $r$ position as a function of $z$.}


\label{fig:EBfield}
\end{figure}

To understand the SMI and proton defocusing, we look at the wakefields that a proton experiences when traversing the plasma. Figure \ref{fig:EBfield}a shows the force - the contributions from the radial electric field $E_r$ and the azimuthal magnetic field $B_\phi$- that affects the proton with the highest radial momentum $p_r$. The resulting force in CGS units is:
\begin{equation}
F_r = E_r - B_{\phi}
\end{equation}
In Figure \ref{fig:EBfield}b we show the radial proton position $r$ as a function of the longitudinal coordinate $z$ along the plasma. 

Figure \ref{fig:EBfield}a shows that the proton deflection results mainly from the radial electric field $E_r$ while the azimuthal magnetic field $B_\phi$ contributes around ten-times less. 
The transverse force acts on the protons from approximately 2 to \SI{3.6}{m} propagation distance in plasma ($\approx 1.6\,\rm{m}$).

We observe that the proton passes through a field that reaches up to 20\% of the maximum wave-breaking field $E_0 = m_e c \omega_p /e$, which corresponds to $\sim 0.5\,\rm{GV/m}$ ($E_0\sim 2.5\,\rm{GV/m}$ for the $n_{pe} = 7 \times 10^{14}$ electrons/cm$^3$). The average experienced wakefield is around $0.15\,E_0 \sim 0.375\,\rm{GV/m}$.
Plasma wakefields in the simulation reach up to $0.4\,E_0 \sim 1\,\rm{GV/m}$, but the proton experiences only a fraction of this field. 

Simulations show that the maximum defocusing angle $\theta$ is on the order of \SI{1}{mrad}. 

In a previous article \cite{Turner2016}, we estimated this angle as:
\begin{equation}
\theta \sim \sqrt{\frac{m_e}{\gamma m_p}}
\label{eq:est}
\end{equation}
where $\gamma$ is the relativistic Lorentz factor of the proton and $m_p$ is the proton mass.
This estimate assumes that the proton experiences a wakefield of $0.5\,E_0$ over a distance of \SI{40}{cm} (see the pink box in Fig. \ref{fig:EBfield}a). The resulting radial kick is approximately $1\,\rm{mrad}$. Now we see that while the average force was overestimated by a factor of four, the interaction distance was underestimated by the same factor, so Eq. \ref{eq:est} is still valid and the maximum defocusing angle estimate stays at $\approx 1\,\rm{mrad}$.

The proton shown in Figure \ref{fig:EBfield} first experiences a focusing force towards the beam axis. There are no transverse wakefields on the axis, the proton crosses the axis. Then the proton enters a defocusing wakefield and experiences a defocusing force away from the beam axis. Our simulations confirmed that this kind of trajectory is typical for strongly defocused protons. 

The wakefields that the proton experiences in Figure \ref{fig:EBfield} shift from the focusing to the defocusing phase of the wake, while the proton's longitudinal position $\xi$ remains the same (\SI{15}{cm} after the seeding laser pulse). The wakefield at this $\xi$-position is determined by the proton distribution at the preceding part of the beam. Since the shape of the proton bunch evolves during SMI, the phase of the wakefield changes as well. The phase shift is an accumulated effect that increases along the proton bunch.
We observed that protons at the rear part of the bunch generally defocus: if they start in the focusing region they appear in the defocusing field as the wakefields shifts. If they start in the defocusing field, they move radially too far out to experience a significant focusing field after the phase shift.
As a consequence, the number of protons in the decelerating focusing phase decreases towards the end of the proton bunch and the wakefield amplitude saturates.

\subsection{Maximum defocusing angle $\theta$}
\begin{figure}[tb!]
		\includegraphics[width = 0.5\textwidth]{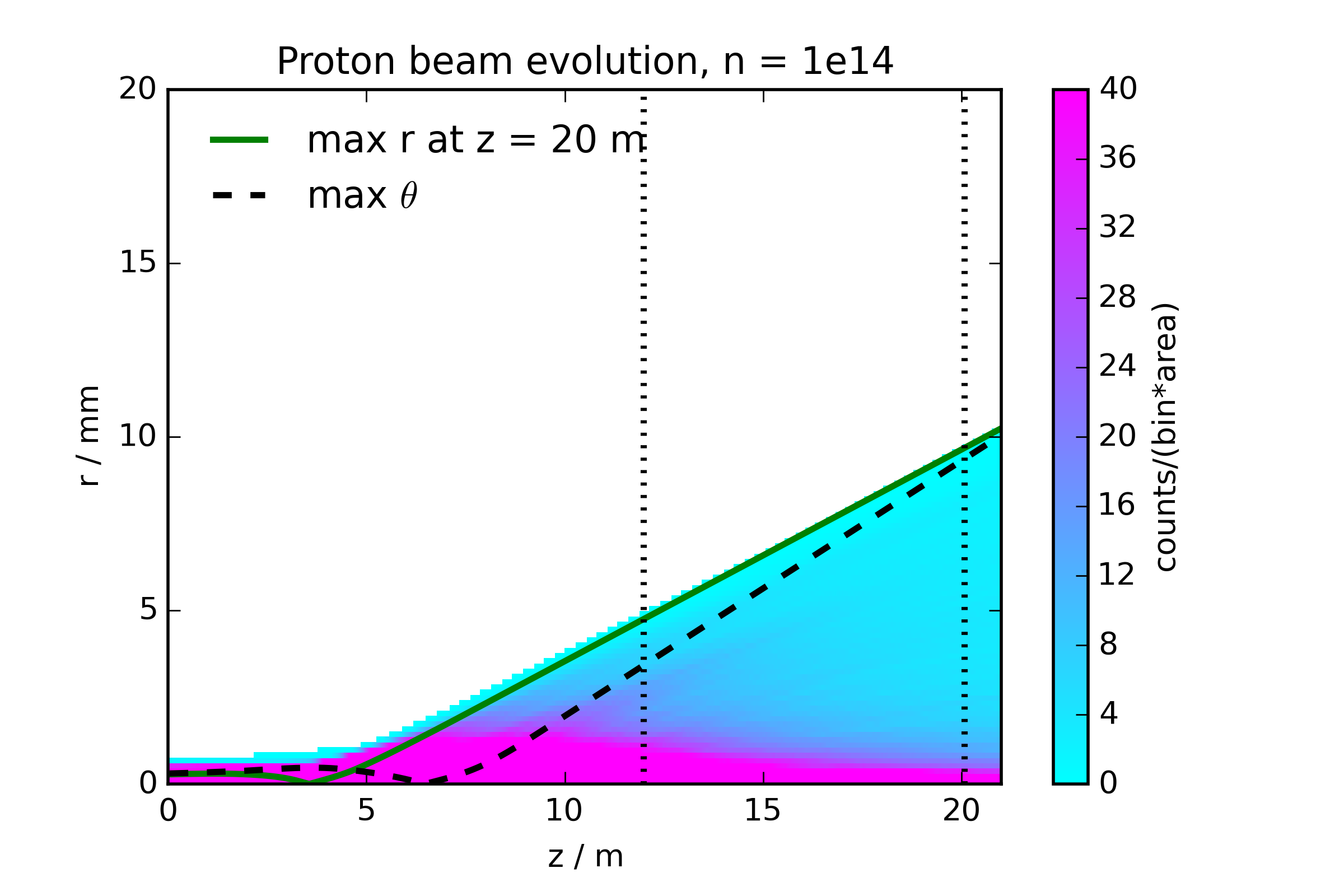}
		\includegraphics[width = 0.5\textwidth]{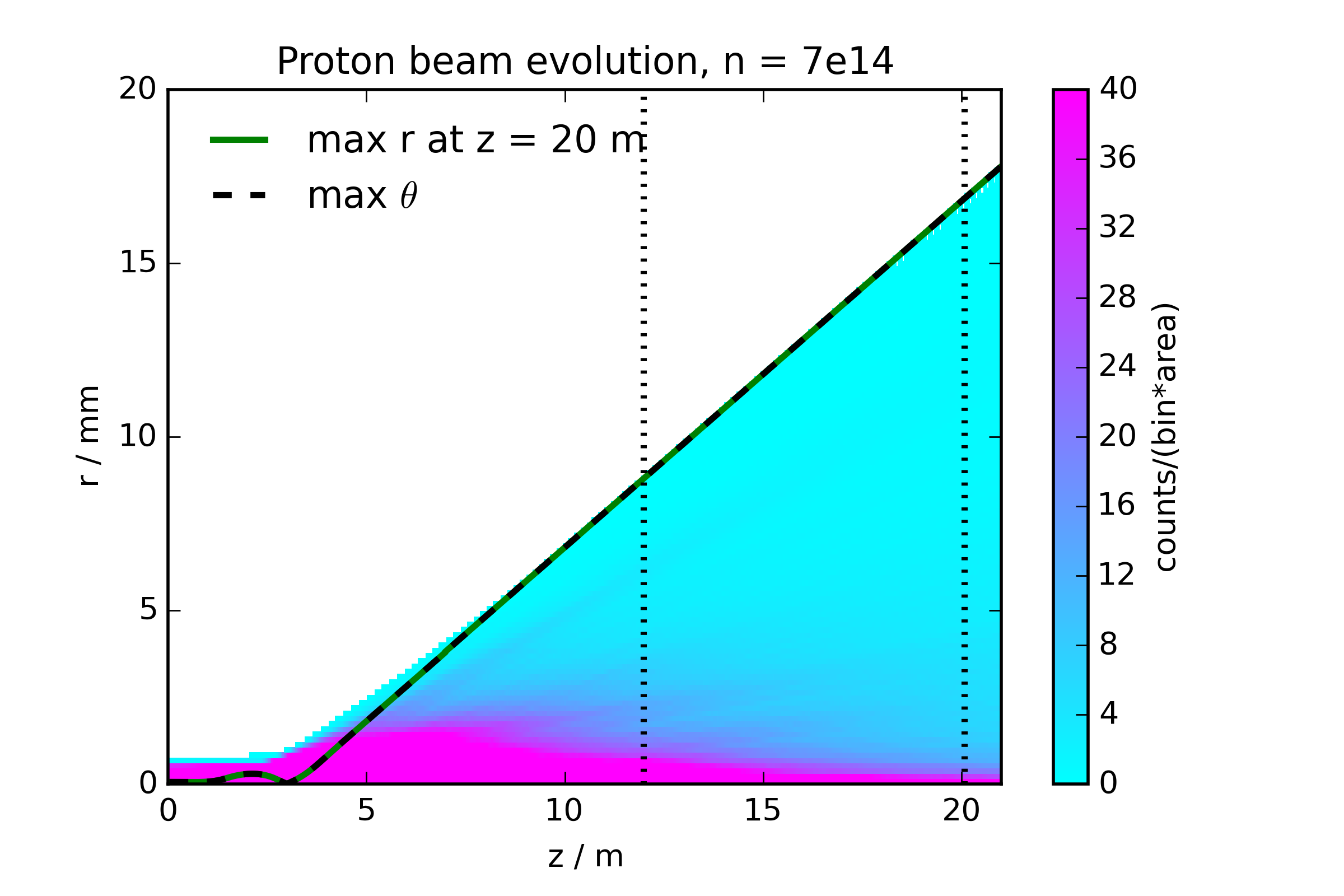}
		\caption{Proton beam density evolution for a plasma density of $n_{pe} = 1 \times 10^{14}\,\rm{electrons/cm}^{-3}$ (top) and $n_{pe} = 7 \times 10^{14}\,\rm{electrons/cm}^{-3}$ (bottom). The plasma starts at $z = 0$ and ends at $z = 10\,\rm{m}$.}
		
		\label{fig:traj}
\end{figure}
We aim to experimentally measure the maximum angle of the defocused protons to indirectly prove that strong plasma wakefields were created. The proton beam size is measured \SI{2}{m} and \SI{10}{m} downstream from the plasma exit by inserting two imaging screens. If protons with the largest radial position at the first screen also have the maximum defocusing angle, we can obtain this angle $\theta$ from the two beam images.

In Figure \ref{fig:traj} we show a density plot of the proton beam evolution in and outside of the plasma (where $z = 0$ corresponds to the entrance of the plasma). In the top plot of Figure \ref{fig:traj} the plasma density is $n_{pe} = 1 \times 10^{14}\,\rm{electrons/cm}^{-3}$ and we see that at $z = 12\,\rm{m}$ the outermost particle (green line) does not correspond to the one with the maximum defocusing angle $\theta$ (black line). In the bottom plot of Figure \ref{fig:traj} though, where the plasma density is $n_{pe} = 7 \times 10^{14}\,\rm{electrons/cm}^{-3}$, we see that the outermost particle is the one with the maximum defocusing angle at both screens.

We analysed the trajectories for plasma densities ranging from $n_{pe} = 1 \times 10^{12}$ to $1.4 \times 10^{15}\,\rm{electrons/cm}^{-3}$ and concluded that we are able to measure the maximum defocusing angle $\theta$ for plasma densities above $n_{pe} = 5 \times 10^{14}\,\rm{electrons/cm}^{-3}$. 
The maximum defocused protons get their radial kick in proximity of the maximum wakefield amplitude along the plasma. The maximum wakefield amplitude is present at the saturation point of the SMI along the plasma. Consequently, for plasma densities above $n_{pe} = 5 \times 10^{14}\,\rm{electrons/cm}^{-3}$ the origin location of the maximum defocused protons and SMI saturation point can be reconstructed by measuring the maximum proton beam size and calculating the maximum defocusing angle $\theta$.

\section{Conclusions}
In AWAKE the main defocusing force that acts on the protons during the development of the SMI results from the radial electric field $E_r$. In the experiment, we expect that plasma wakefields on the order of \SI{0.4}{GV/m} defocus protons over a distance of $\Delta z \approx 1.6\,\rm{m}$. The resulting maximum proton defocusing angle is $\approx\,$\SI{1}{mrad}. Consequently, measuring defocusing angles around \SI{1}{mrad} proves that strong plasma wakefields were created and indirectly confirms the development of SMI. 

We showed that the two-screen diagnostics can measure the maximum defocusing angle $\theta$ for plasma densities higher than $n_{pe} = 5 \times 10^{14}\,\rm{electrons/cm}^{-3}$. Having the maximum defocusing angle $\theta$ and the radial proton beam size, we can determine the saturation point of the SMI along the plasma.

\null
\end{document}